\documentstyle[12pt]{article}
\begin{document}

\baselineskip 24pt

\newcommand {\sheptitle}{Avoiding the theorem of Lerche and Shore}

\newcommand{\shepauthor}{K.J. Barnes}

\newcommand{\shepaddress}{Department of Physics and Astronomy,\\
University of Southampton\\
Southampton, SO17 1BJ\\
United Kingdom\\
\smallskip
Telephone: +44 1703 592097\\
Email: kjb@phys.soton.ac.uk}

\newcommand{\shepabstract}
{\noindent {\bf Abstract} \hfill \\
\noindent Supersymmetric $\sigma$-models obtained by constraining 
linear
supersymmetric field theories are ill defined.
Well defined subsectors parametrising Kahler manifolds
exist but are not believed to arise directly from constrained linear 
ones.
A counterexample is offered using improved understanding of
membranes in superstring theories leading to crucial central terms
modifying the algebra of supercharge densities.}

\begin{titlepage}
\vspace{.4in}
\begin{center}
{\large{\bf \sheptitle}}
\bigskip \\
\shepauthor \\
\mbox{ } \\
{\shepaddress}
\end{center}
\vspace{.5in}
{\shepabstract}\\
\vfill
\end{titlepage}
Massless Goldstone bosons arise from components of global symmetries 
which are spontaneously broken. There is no extra symmetry for 
Goldstone bosons in supersymmetr
Instead the supersymmetry forces complexification of scalars.
This leads to an increased number of massless excitations in 
general,
with complete doubling of the original number in some cases.
The special cases when the coset space manifold, $G/H$, of the
original Goldstone bosons is Kahler might be expected to be an
exception in view of the seminal work of Zumino \cite{1}.
This does indeed confirm that a non-linear supersymmetry model
can be established without any increase in the number of Goldstone 
bosons.
However the theorem frequently attributed to Lerche and Shore,
following early work by Ong \cite{2}, appears to prove that such
models can never result from constraining linear supersymmetric 
ones.
The formal proofs in Lerche \cite{3}, and in Kotcheff and Shore
\cite{4}, reveal a striking similarity to the work of
Witten \cite{5}, in which the impossibility of partially
breaking extended global supersymmetries $(N > 1 )$ to lower values 
was proposed.
Indeed it was this similarity which prompted the current work.
Once Bagger and Wess \cite{6}, and subsequently
Hughes, Liu and Polchinski \cite{7} had produced
(non--linear) counter examples to the Witten analysis
it seemed likely that the Lerche and Shore proofs would also fail,
The key contribution could be argued to be that of
Hughes and Polchinski \cite{8}, which revealed that the
original anticommutator algebra for supersymmetric charges
had to be generalised to include a central term at the
underlying current density level.
They attributed this revision to the more
modern viewpoint that supermembranes are just as
fundamental as elementary particles in string theory.

This reinterpretation is the current starting point.
The generalisation of
\begin{equation}
\left\{ Q_{A \alpha}, Q_{B \dot{\beta }} \right\} =
2 ( \sigma^\mu )_{\alpha \dot{\beta}} \delta_{A B}
\end{equation}
to local form is

\begin{equation}
	\partial_\mu T \left( j_{A \alpha}^\mu (x)
\bar{j}_{B \dot{\beta}}^\nu (y) \right) 
	=  2 \left( \sigma^\rho \right)_{\alpha \dot{\beta}} T_\rho^\nu
\delta^4 (x - y) \delta_{AB} 
	+  2 \left( \sigma^\nu \right)_{\alpha \dot{\beta}} C_{AB}
\delta^4 ( x - y) ,
\end{equation}
where Schwinger terms which are irrelevant to this analysis are 
ignored \cite{8}.
The key feature is provided by the central terms
$C_{AB}$ which give infinities of unclear covariance on integration
over fixed volume.
Possibly this was why this was previously overlooked.
One might wonder if equation (2) could be restricted by the fact
that the Hughes and Polchinski treatment was in two dimensions.
But they appear to be taking advantage of the fact that
$T^{\mu \nu}$ is not the unique conserved symmetric tensor since
$T^{\mu \nu} + C \eta^{\mu \nu}$ is also conserved.
This does not depend on being in two or fewer dimensions.
It seems that this is one of those situations where the symmetry of 
the hamiltonian is
larger than the symmetry of the $S$-matrix.
At any rate equation  (2) is clearly finite and Lorentz invariant.
From it follow the usual consequences of degenerate multiplets
for unbroken supersymmetries and Goldstone fermions for those
that are broken.
In momentum space, with $C_{AB}$ diagonal and $<T^{\mu \nu}> =
\Lambda \eta^{\mu \nu}$, we have
\begin{equation} q_\mu < j_{A \alpha}^\mu (q) \bar{j}_{A 
\dot{\beta}}^\nu >
= 2 \left( \sigma^\nu \right)_{\alpha \dot{\beta}}
(\Lambda + C_{AA} ) + 0 (q)
\end{equation}
where there is no sum over $A$.
For those $A$ such that $\Lambda + C_{AA} \neq 0$,
equation (3) implies a
$1/\! \not{\!q} $ singularity in the two current correlations,
$J_{A \alpha}^\mu$
couples the vacuum to a massless fermion with coupling
strength $[ 2 ( \Lambda + C_{AA}) ]^{1/2} $.
It also follows that $\Lambda + C_{AA} \geq 0$.
It should now be obvious how to evade the extra unwanted Goldstone 
bosons,
in the case where the underlying coset manifold is indeed Kahler.
The classical analyses of Coleman, Wess and Zumino \cite{9}, and
Callan, Coleman, Wess and Zumino \cite{10}
were extended in the case of non-linearly realised supersymmetry
by Volkov and Akulov \cite{11}.
This paper follows the elegant treatment of Itoh, Kugo and Kunitoma
\cite{12} based upon the
very complete generalisations of the classical analyses by
Bando, Kuramoto, Maskawa and Uehara \cite{13}, and the same
authors in \cite{14} and \cite{15}.
Finally, we bring attention to the further clarifications made by
Volkov \cite{16} and so elegantly presented by Ogievetsky 
\cite{17}.

The crucial point of extending the underlying algebra of 
supercharge
current densities by central terms, has to be combined not merely 
with a
Kahler $G/H$, but that manifold has to be re-expressed as a quotient 
space of the complexified $G$
(usually called $G^c$ ) by a maximially
extended complex extension of $H$ (usually called $\hat H$).
In this treatment this will appear as an explicit mapping
manifesting the homeomorphism between $G/H$ and $G^c / \hat H$.

A concrete example is offered in the form of the simplest possible 
case of
$G/H = SU_2 / U_1$, usually
called the complex projective space $CP2$, although it is
a straightforward task to extend to all similar (i.e. Kahler) but
more complicated cases.
The starting point is a recent, interesting but incomplete, attempt 
to
generalise the ideas of chiral perturbation theory to the
supersymmetric level by Barnes, Ross and Simmons \cite{18}.
It is instructive to see how the ambiguities arise in this chiral
$SU_2 \times SU_2 $ based
model and we adapt the notation of the original only slightly.
The original (unconstrained) supersymmetric action is constructed 
from
four (complex) chiral superfields. In components, with
\begin{equation}y^m = x^m + i \theta \sigma^m \bar{\theta} , 
\end{equation}
these have the form
\begin{eqnarray}
\Phi ( x , \theta \bar{\theta} ) &=& \phi (y) + \sqrt{2} \theta 
\lambda_{\phi} (y) +
\theta^2 F_{\phi} (y) , \\
\Sigma_3 ( x, \theta , \bar{\theta} ) & = & \sigma_3 (y) + \sqrt{2} 
\theta \lambda_3 (y) + \theta^2 F_\sigma (y)
,\\
\Pi_A ( x , \theta , \bar{\theta} ) &= & \pi_A (y) + \sqrt{2} 
\lambda_A (y) + \theta^2 F_A (y) ,
\end{eqnarray}
where $\sigma^m = (-1, \tau^a)$, and the $\tau^a$ are the Pauli 
matrices
$(a = 1, 2, 3).$
The chiral superfields $\Pi_A (A = 1, 2)$, and $\Sigma_3$ transform 
as a
triplet under $SU_2$, where the third direction
is that of the intended spontaneous breaking, and $\Phi$ is a 
singlet.
It has previously been noted by Barnes, Generowicz and Grimshare
\cite{19} that the chiral $SU_2$ generated by the first two 
components
of the axial generators together with the third component of the 
vector
generators leads indeed to a Kahler manifold of the type
$SU_2/U_1$.
This is embedded in the chiral $\frac{SU_2 \times SU_2}{SU_2}$ 
structure
exactly so as to give the $\pi_A$ pseudoscalar nature in their real
parts, and correspondingly $\phi$ and $\sigma_3$ scalar nature.
The most general supersymmetric action is then written as
\begin{equation}
 I = \int d^8 z ( \bar{\Phi} \Phi + \bar{\Sigma}_3 \Sigma_3 +
\bar{\Pi}^A \Pi_A) 
+ \int d^6s W + \int d^6 \bar{s} \overline{W} , 
 \end{equation}
where the superpotential $W$ is a functional of chiral superfields 
only.
Combining the $\Sigma_3$ and $\Pi_A$ fields into the
matrix
\begin{equation}M = \Sigma_3 \tau^3 + \Pi_A \tau^A , \end{equation}
where the chiral $\gamma_5$ factors are now suppressed,
reveals that, under chiral $SU_2 \times SU_2$, M transforms as
\begin{equation}M \to L MR^\dag , \end{equation}
and taking
\begin{equation}W = k ( \det M + f_\pi^2 ) \Phi , \end{equation}
where $k$ is a constant,
ensures that the model reduces to the usual bosonic chiral
model below the supersymmetry breaking scale
provided that $f_\pi$ is required to be real.
Notice that $\sigma^2$ of reference \cite{18} now appears in the
guise of $(- \sigma_3)^2$.
The advantage of this change of notation will become clear
later.
This starting action now yields the potential
\begin{equation}
\begin{array}{ll}
V & = F_\sigma \bar{F}_\sigma + F_A \bar{F}_A + F_{\phi} \bar{F}_\phi 
\\
 & = 4k^2 \phi \bar{\phi} ( \sigma_3 \bar{\sigma}_3 + \pi_A 
\bar{\pi}_A ) \\
 & + k^2 [ f_\pi^2 - \sigma_3^2 - \pi_A \pi_A ] [ f_\pi^2 - 
\bar{\sigma}_3^2 -
\bar{\pi}_A \bar{\pi}_A] . \end{array} \end{equation}
The minimum of this potential is clearly $V = 0$ which
may be achieved by giving the fields the following $SU_2 \times SU_2$ 
breaking
vacuum expectation values (VEVs)
\begin{equation}<\sigma_3 > = f_\pi \; \; \; <\pi_A> = 0 = < \phi > 
.
\end{equation}

Importantly no auxiliary field acquires a VEV with these 
assignments
so supersymmetry is manifestly not broken in this model.
The formal limit $k \to \infty$ leaves the action
\begin{equation}I = \int d^8 z ( \Sigma_3 \bar{\Sigma}_3 +
\Pi_A \bar{\Pi}_A + \Phi \bar{\Phi} ) ,
\end{equation}
with the superfields subject to the constraint
\begin{equation}\Sigma_3^2 + \Pi_A \Pi_A = f_\pi^2 ,
\end{equation}
with the consequence that the superfield $\Phi$ takes no part in the 
interactions and can
be ignored as spectator field.
Eliminating $\sigma_3$, $\lambda_3$ and $F_3$ by substituting the 
constraints into the
kinetic part of the Lagrangian to obtain the leading term in the low 
momentum expansion
(each fermion is considered to have associated with it a factor
of the square root of the momentum scale) gives exactly the 
non-linear
(Zumino) Lagrangian as reported in reference \cite{18}.
As stated there, the interaction terms involving pseudo-Goldstone 
bosons and, or,
the fermionic superpartners are not uniquely specified.
However, the structure of the non-linear Lagrangian describing the 
Goldston pions alone
(when the scalar fields are taken to be real, and the fermions
supressed), is quite independent of the structure in which it is
is now embedded.
This applies also to the Kahler subset of fields, but in the previous 
conventional wisdom
this subsector alone was prohibited from arising in this manner by
the theorem of Lerche and Shore.
Finally, it is necessary to express the manifold in the
$G^c/\hat{H}$ complex form.
The key is to introduce the projection operator $\eta$
with the properties $\eta^2 = \eta $ and $\eta^\dag = \eta $ which 
is
made possible because of the change in the underlying algebra
of supercharge densities with the central terms.
This can be taken, in this notation, to be
\begin{equation} \eta = \frac{1 + \tau_3}{2} , 
\end{equation}
and it is trivial to confirm that the property
\begin{equation} \hat{h} \eta = \eta \hat{h} \eta , 
\end{equation}
picks out $\tau^+ $ and $\tau^3$ as the four members of
the complex subgroup.
(There is a two way alternative choice
at this point, but this has become standard.)
With this in mind rewriting $M$ as
\begin{equation}
M = \Sigma_3 \tau^3 + \frac{i \Delta \tau^3}{2} - \frac{i \Gamma 
\tau^-}{2}
, 
\end{equation}
so that, now leaving out the $f_\pi^2$ terms,
\begin{equation}
I = \int d^8 z ( \Sigma_3 \bar{\Sigma}_3 + \frac{\Gamma 
\bar{\Gamma}}{4} +
\frac{\Delta \bar{\Delta}}{4} + \Phi \bar{\Phi} ) , 
\end{equation}
and
\begin{equation}
\begin{array}{ll}
V & =  	4 k^2 \phi \bar{\phi} \left(\sigma_3 \bar{\sigma}_3 +
	\gamma \bar{\gamma} + \delta \bar{\delta} \right)\\
  & +  	k^2 \left[ \sigma_3^2 + \frac{\gamma \delta}{4} \right]
	\left[ \bar{\sigma}_3^2 + \frac{\bar {\gamma} \bar{\delta}}{4}
\right] . \label{20}
 \end{array} .
 \end{equation}
 In the formal limit as $k \to \infty$, the action becomes
 \begin{equation}
 I = \int d^8 z \frac{\Gamma \bar{\Gamma}}{4} , \end{equation}
as the constraints are satisfied by the superfield conditions
\begin{equation} \Sigma_3 = 0 \;\; \; {\rm and} \; \; \; \Delta = 0 . 
\end{equation}
The superfield $\Phi$ can again be ignored as a non-interacting 
spectator.
Notice that the single complex superfield $\Gamma$ is all that 
remains in
the action, and it is not constrained.

To describe the coset space of the chiral sphere \cite{19}, the 
real
part of $\pi^A $ is written as $M^A$, and so
\begin{equation}
L = \exp \left\{ \frac{-i}{2} \theta( \phi) \frac{M_A \tau^A}{\phi} 
\right\} ,
\end{equation}
where $\theta(\phi)$ is any arbitrary function of
\begin{equation} \phi = [ M_A M^A ]^{1/2} , \end{equation}
divided by the pion decay constant $f_\pi$, and where the chiral
$\gamma_5$ dependence is again suppressed.
The arbitrariness may be viewed as the
freedom to change coordinate systems on the surface of the sphere, 
or
to redefine the field variables describing the pions.
Now this can alternatively be written in the form \cite{12}
\begin{equation}
L = \exp \left( \frac{- i \gamma \tau^-}{2} \right)
\exp
\left( \frac{ -i \delta \tau^+}{2} \right)
\exp \left( \frac{ - V \tau^3}{2} \right) ,
\end{equation}
using the complex subgroup $\hat{H}$. Moreover the expression
\[ \exp \left( \frac{ - i \gamma \tau^-}{2} \right) \]
gives the explicit mapping of the homeomorphism
between $G/H$ and $G^c/ \hat{H}$.
In the general coordinate system 
\begin{equation}
\frac{ \gamma \bar{\gamma}}{4} =
\tan^2 \left( \frac{ \theta}{2} \right) , \end{equation}
and it is known from reference \cite{12} that the Kahler potential is 
given by
\begin{equation}
K = ln {\det}_\eta \left\{
	\exp \left( \frac{i \bar{\gamma} \tau^+}{2} \right)
	\exp \left( \frac{ - i \gamma \tau^-}{2} \right) \right\} ,
	\end{equation}
where the notation indicates that the determinant is to be
taken in the top left hand corner of the matrix in this 
representation.
This reveals at once that
\begin{equation}
K = \ln \left[ 1 + \frac{ \bar{\gamma} \gamma}{2} \right] 
 = \ln \left[ sec^2 \left( \frac{\theta}{2} \right) \right] = V
 \end{equation}
which is the desired result.
Note that, although the general coordinate notation is most
convenient in this context, reliance is placed on the results of 
reference
\cite{12}.
The Kahler nature of
the potential is demonstrated by diagonalizing the metric in 
stereographic
coordinates, $z$ and $\bar{z}$,
and revealing the holomorphic nature of the transformations in the
usual manner.

Although this demonstration that Kahler potentials can arise
from constrained linear supersymmetric schemes has used only $CP2$,
there seems no reason whatsoever why this can not be generalised
directly to larger
Kahler manifolds -- in particular to $CPN$.

The author is grateful to Professor D A Ross for
raising his interest in this type of work. This work is
partly supported by PPARC grant number GR/L56329.

\end{document}